\documentclass[10pt]{iopart}
\eqnobysec

\def\qed{\vrule height8pt width6pt depth1pt}

\def\C{\bar\Gamma}
\def\Q{{\cal Q}}
\def\T{{\rm T}}
\def\X{{\rm X}}
\def\K{K}
\def\P{{\cal P}}

\def\proof{\bigskip\noindent{\bf Proof:}}
\begin{document}

\title{Observables for the polarized Gowdy model}

\author{C G Torre}

\address{Department of Physics, Utah State University, Logan, UT 84322-4415, USA}

\begin{abstract}
We give an explicit characterization of all functions on the phase space for the polarized Gowdy 3-torus spacetimes which have weakly vanishing Poisson brackets with the Hamiltonian and momentum constraint functions.

\end{abstract}

\section{Introduction}

In the canonical description of gravitation it is desirable for certain purposes ({\it e.g.}, isolation of the degrees of freedom of the gravitational field \cite{FM1979}, canonical quantization \cite{Rovelli2000}, {\it etc.}) to have mathematical control over a suitable number of  ``gauge invariant'' dynamical variables --- the so-called {\it observables} of the theory. More precisely, following the general theory of constrained Hamiltonian systems \cite{HT1992}, the observables are defined as functions on the gravitational phase space which have (weakly) vanishing Poisson brackets with the (Hamiltonian and momentum) constraint functions.\footnote{In the past these particular quantities have been called ``Dirac observables''. In Rovelli's terminology they are known as ``complete observables'' \cite{Rovelli2002}. Kucha\v r calls these quantities ``perennials'' \cite{Kuchar1993}. For the purposes of this paper we shall simply call them ``observables''.} Finding all the observables is tantamount to characterizing the ``reduced phase space'', which is equivalent to the set of diffeomorphism equivalence classes of solutions to the field equations \cite{FM1979}. 

  Finding observables for generic vacuum spacetimes has proved to be a very difficult task --- finding all of the observables is roughly equal in difficulty to finding all solutions to the field equations --- and only a handful of observables are known explicitly.  There are restricted classes of spacetimes --- usually arising via the imposition of some spacetime symmetry --- where one can get a mathematically explicit description of the observables. Normally these spacetimes are described by Hamiltonian systems with a finite number of degrees of freedom (see, {\it e.g.}, \cite{ATU1993}), but there are a couple of systems with infinitely many degrees of freedom where the observables have been explicitly computed \cite{Neville1993, Torre1991}.

Here we shall give an explicit characterization of all observables for the polarized Gowdy spacetimes. Gowdy spacetimes have compact spatial slices --- here assumed to have the topology of a 3-torus --- and admit a two-dimensional, orthogonally transitive, Abelian isometry group with spacelike orbits \cite{Gowdy1974}. When the Killing vector fields generating the isometry group are each hypersurface orthogonal the model is ``polarized''.  (In the sequel we shall simply refer to the model being considered as the ``Gowdy model''.) Physically, these spacetimes are interesting because they  correspond to inhomogeneous cosmologies and provide a non-trivial quantum gravity model (see, {\it e.g.}, \cite{Torre2002} and references therein). From a more mathematical point of view, these spacetimes are particularly interesting because they are described by a dynamical system which possesses infinitely many degrees of freedom. Moreover, the isolation of the observables in the Gowdy model turns out to be somewhat intricate compared to the models studied in \cite{Neville1993, Torre1991} because of the compact topology of the spatial hypersurfaces. 

The principal tool used here to find the observables is a set of adapted canonical coordinates for the Gowdy phase space found in \cite{RT1996}. These variables, which are analogous to those introduced by Kucha\v r for Einstein-Rosen spacetimes \cite{Kuchar1971}, show that the Gowdy model can be viewed, for the most part, as a parametrized scalar field theory \cite{Dirac1964, Kuchar1976} on a fixed two-dimensional background spacetime -- much as occurs in the Einstein-Rosen model. However there are some additional features brought into play by the compact spatial slices. Firstly, besides the scalar field degrees of freedom there is an additional single degree of freedom built, roughly speaking, from the constant part of the metric for the spatial metric and its conjugate momentum.  Secondly, there is an additional quadratic constraint on the scalar field degrees of freedom which is responsible for a conical singularity in the constraint surface in phase space. 

In terms of the adapted canonical variables the Gowdy model observables mainly represent the field degrees of freedom (modulo the transformation group generated by the constraint just mentioned) and are constructed as correlations between the canonical variables representing the scalar field and other canonical variables representing embeddings of spatial slices into the background spacetime. These correlations are simply the ``many-fingered time'' evolution of the field variables dictated by the symmetry-reduced Einstein equations. The additional degree of freedom mentioned above can be described by a canonical pair of variables neither of which appear in the Hamiltonian so they are constants of the motion and hence observables. 

In \S2 we define the Gowdy model as a constrained Hamiltonian system and give a precise definition of the observables. In \S3 we give a brief characterization of the reduced phase space in a standard gauge. In \S4 we review the construction of the canonical variables in \cite{RT1996} which facilitate the construction of the observables. Our main result --- an explicit characterization of the observables in a chart on phase space --- is obtained via a 2 step procedure in \S 5.

\section{The model}

The Gowdy model we will study arises by assuming spacetime is not flat, that its manifold is ${\cal M}={\bf R}^+\times {\bf  T}^3$ with spacelike ${\bf  T}^3$, and that there is an Abelian 2-parameter isometry group, $G={\bf T}^2$, with spacelike  orbits ${\bf T}^2\subset {\bf T}^3$ generated by a pair of commuting, hypersurface-orthogonal Killing vector fields. The symmetry-reduced theory lives on the space of orbits $M={\cal M}/G$, which is a cylinder, $M={\bf R}^+\times {\bf S}^1$.  We will use coordinates $(t,x,y,z)$ on $\cal M$, where $t>0$, and $(x,y,z)\in (0,2\pi)$. These coordinates are chosen such that the $t=const.$ surfaces foliate $\cal M$ with spacelike ${\bf T}^3$ hypersurfaces (coordinates $(x,y,z)$), and such that the Killing vector fields are ${\partial\over\partial y}$ and ${\partial\over\partial z}$.  On $M$, $t$ is a time coordinate labeling a spacelike foliation by circles for which $x$ is a coordinate. 
In such coordinates
the metric $g$ on $\cal M$ can be put into the form \cite{RT1996}
\begin{equation}
\eqalign{
g=[&-(N^{\perp})^2 + e^{\gamma-\psi}(N^x)^2]dt^2
+ 2e^{\gamma-\psi}N^xdtdx
+ e^{\gamma-\psi}dx^2\cr
&+ \tau^2e^{-\psi}dy^2 + e^\psi dz^2.}
\end{equation}
The functions $N^\perp(t,x)$ and $N^x(t,x)$ determine the lapse function and shift vector for $G$-invariant spacelike foliations of $\cal M$ or, equivalently, the lapse and shift for spacelike foliations of $M$. The function $\tau(t,x)>0$ defines the area of the orbits of $G$ and is often used as a time coordinate. The gradient of $\tau$ is assumed to be everywhere timelike.
  The functions $\gamma(t,x)$ and $\psi(t,x)$ are unrestricted.

A Hamiltonian formulation of the symmetry-reduced Einstein equations is given in \cite{RT1996}, and we shall use this formulation here.  Using a prime and dot to denote differentiation with respect to $x$ and $t$ respectively, the canonical action functional for the theory defined on $M$ is given by
\begin{equation}
\eqalign{
S&[N^\perp,N^x,\gamma,\tau,\psi,\Pi_\gamma,\Pi_\tau,\Pi_\psi] =\int_{t_1}^{t_2}dt\,\int_0^{2\pi}
dx\,\Bigg(\Pi_\gamma\dot\gamma +\Pi_\tau\dot \tau + \Pi_\psi\dot\psi 
\cr
 &-N^\perp e^{(\psi-\gamma)/2}\Big[-
\Pi_\gamma\Pi_\tau+2\tau^{\prime\prime}-
\tau^\prime\gamma^\prime
+{1\over2}\left(\tau^{-1}\Pi_{\psi}^2+\tau\psi^{\prime2}\right)
\Big]\cr
&- N^x\Big[
-2\Pi_\gamma^\prime
+\Pi_\gamma\gamma^\prime+\Pi_\tau\tau^\prime+\Pi_\psi\psi^\prime\Big]\Bigg),}
\end{equation}
which clearly displays the canonical coordinates\footnote{In what follows we shall always denote the momentum conjugate to a variable, say, $\chi$, by $\Pi_\chi$.} $(\gamma, \tau, \psi, \Pi_\gamma, \Pi_\tau, \Pi_\psi)$ for the phase space $\Gamma$, and the Hamiltonian and momentum constraint functions:
\begin{equation}
H_\perp = e^{(\psi-\gamma)/2}\Big[-
\Pi_\gamma\Pi_\tau+2\tau^{\prime\prime}-
\tau^\prime\gamma^\prime
+{1\over2}\left(\tau^{-1}\Pi_{\psi}^2+\tau\psi^{\prime2}\right)\Big],
\label{Hperp}
\end{equation}
\begin{equation}
H_x=-2\Pi_\gamma^\prime
+\Pi_\gamma\gamma^\prime+\Pi_\tau\tau^\prime+\Pi_\psi\psi^\prime.
\label{Hx}
\end{equation}
The constraints $H_\perp=0=H_x$, which are obtained upon variation of the action with respect to the lapse and shift functions $(N^\perp, N^x)$, define the {\it constraint surface} $\C\subset\Gamma$.

The restriction that the spacetime gradient of $\tau$ is timelike leads to a corresponding restriction on the phase space $\Gamma$:
\begin{equation}
\Pi_\gamma <-|\tau^\prime|.
\end{equation}
Strictly speaking, this phase space restriction is theory-dependent in the sense that it depends upon the form of the action, but we will view it as an {\it a priori} feature of the phase space $\Gamma$.

With the Hamiltonian formulation of the model in hand we can define the observables as follows.

\bigskip
\noindent
{\bf Definition.}
{\it {\rm Observables} are functions on the polarized Gowdy phase space $\Gamma$,
\begin{equation}
F = F(\tau, \gamma,\psi,\Pi_\tau,\Pi_\gamma,\Pi_\psi),
\end{equation}
which have weakly vanishing Poisson brackets with the constraint functions:
\begin{equation}
[F,H_\perp] = 0=[F,H_x],\quad {\rm modulo}\ H_\perp,\, H_x.
\end{equation}
An observable is {\rm trivial} if it vanishes on $\C$. Two observables $F_1$ and $F_2$ are {\rm equivalent} if $F_1-F_2$ is  trivial.}
\bigskip

The observables correspond to functions on the {\it reduced phase space} $\hat \Gamma$. This space is obtained from the original phase space $\Gamma$ as follows. The constraint functions generate a set of canonical transformations on $\Gamma$. Because the constraints are ``first class'', this set of transformations restricts to act on the constraint surface $\bar\Gamma$. The reduced phase space $\hat \Gamma$ is the space of orbits of these canonical transformations in $\bar\Gamma$.  We shall refer to the process of restricting to the constraint surface and forming the space of orbits of the canonical transformations on the constraint surface as {\it reduction} by the constraints. (In this case the constraints are $H_\perp=0=H_x$.) There is a 1-1 correspondence between inequivalent observables and functions on the reduced phase space. In particular, if we denote by $\pi\colon\bar\Gamma\to\hat\Gamma$ the projection map from the constraint surface to the reduced phase space, this correspondence is simply pull-back by $\pi$ of functions on $\hat\Gamma$ to observables on $\bar\Gamma$.

\section{Gauge fixed representation of the reduced phase space}

For comparison with our subsequent results  we briefly review a gauge-dependent characterization of the reduced phase space for the Gowdy model being considered here (see, {\it e.g.}, \cite{Pierri2002}).    A convenient set of gauge conditions is provided by
\begin{equation}
\tau = t,\quad \Pi_\gamma = \Pi_0,
\label{gauge1}
\end{equation}
where
\begin{equation}
\Pi_0 = {1\over 2\pi}\int_0^{2\pi} dx\, \Pi_\gamma <0.
\end{equation}
In spacetime terms, these conditions use the area time coordinate --- which is a harmonic coordinate --- and a harmonic conjugate to label points in $M$. Granted these gauge constraints, the Hamiltonian and momentum constraints are equivalent, respectively, to
\begin{equation}
\Pi_\tau = {1\over 2\Pi_0}\left({1\over t}\Pi_\psi^2 + t\psi^{\prime 2}\right)
\label{pitau}
\end{equation}
\begin{equation}
\gamma^\prime =- {1\over\Pi_0} \Pi_\psi\psi^\prime.
\label{gammaprime}
\end{equation}
The constraints (\ref{gauge1}),  (\ref{pitau}) and (\ref{gammaprime}) would define a representation of the reduced phase space as a submanifold in $\Gamma$ were it not for the fact that (\ref{gammaprime}) has a solution for $\gamma$ if and only if 
\begin{equation}
\K:=\int_0^{2\pi}dx\, \Pi_\psi\psi^\prime = 0.
\label{K}
\end{equation}
This quadratic condition is a residual first-class constraint from (\ref{Hperp})--(\ref{Hx}) left untouched by the gauge conditions (\ref{gauge1}).  The canonical transformation generated by (\ref{K}) corresponds to a translation of the harmonic coordinate conjugate to $\tau$. The existence of this residual first-class constraint stems from the fact that the conditions (\ref{gauge1}) only fix coordinates on $M$ up to this translation. 

In this gauge the reduced phase space can be represented as follows. Denote by $\tilde\Gamma$ the space of canonical pairs $(\gamma_0, \Pi_0)$ and $(\psi, \Pi_\psi)$,
where
\begin{equation}
\gamma_0 = \int_0^{2\pi} dx\, \gamma(x).
\end{equation}
$\tilde\Gamma$ is the phase space resulting from the (partial) gauge fixing. $K$ generates a 1-parameter group of canonical transformations on $\tilde\Gamma$:
\begin{equation}
\eqalign{
&\gamma_0\to\gamma_0,\quad \Pi_0\to\Pi_0,\cr
 &\psi(x)\to \psi(x+c),\quad \Pi_\psi(x) \to \Pi_\psi(x+c),
 }
 \label{Kaction}
\end{equation}
where $c$ is a constant and we make the identification $x\sim x+2\pi$.   This transformation group restricts to act on the surface $K=0$. The space of orbits of the group action (\ref{Kaction}) on this surface  can be identified with $\hat\Gamma$. In \S5 we shall give a more explicit -- and gauge invariant -- characterization of the reduced phase space.

The constraint (\ref{K}) also reveals the existence of a conical singularity in the constraint surface at $\psi^\prime=0=\Pi_\psi^\prime$, which is known to exist on general grounds \cite{AMM1982}. The set of phase space points for which $\psi^\prime=0=\Pi_\psi^\prime$ are fixed points of the transformation generated by (\ref{K}) and   represent initial data for homogeneous spacetimes. 

\section{Extended phase space and a change of variables}

Following \cite{RT1996}, it will be convenient to enlarge the phase space slightly and make a change of canonical coordinates before constructing the observables. This enlargement is an instance of a general procedure outlined in \cite{AMM1982} for resolving singularities in the constraint surface, such as we have here.  Because of these singularities some such modification is needed if one is to view canonical general relativity in terms of parametrized field theory \cite{Torre1992}.

Consider the symplectic manifold $T^*S^1$, with canonical coordinates $(\theta,\Pi_\theta)$, where $\theta\in (0,2\pi)$ is a coordinate on the circle and $\Pi_\theta$ is its conjugate momentum.  (The significance of $S^1$ is as the stabilizer group of the singular points in the constraint surface.) Define the extended phase space as
\begin{equation}
\Gamma^\star = \Gamma \times T^*S^1.
\end{equation}
 Next, introduce the (first-class) constraint
\begin{equation}
\Pi_\theta=0,
\end{equation}
which renders $\theta$ ``pure gauge''. Reduction of $\Gamma^\star$ by this constraint yields the original Gowdy phase space $\Gamma$ described above.
The utility of this extension of the phase space can be seen in the context of the canonical variables introduced in \cite{RT1996}.  These are constructed as follows.

 Introduce a normalized measure $\mu$ on the spatial circles $S^1$ embedded in $M$ (not the same $S^1$ as just introduced above!), 
\begin{equation}
\int_0^{2\pi} dx\, \mu(x) = 1.
\end{equation}
Now make the canonical change of variables (on $\Gamma^\star$)
\begin{equation}
 (\tau,\gamma,
\psi,\theta, \Pi_\tau,\Pi_\gamma, \Pi_\psi,
\Pi_\theta) \longrightarrow (T,X,\phi,{\cal Q},
\Pi_T,\Pi_X,\Pi_\phi, \Pi_{\cal Q})
\label{CT1}
\end{equation}
given by 
 \numparts
\begin{eqnarray} 
 T&=
-{1\over\Pi_0}\tau,\\
\Pi_T&=-\Pi_0\left(\Pi_\tau + \left[\ln\left({ \Pi_\gamma-
\tau^\prime\over\tau^\prime+\Pi_\gamma}\right)\right]^\prime\right),
\\
X(x)&=\theta +
\int_0^{2\pi}dx^{\prime\prime}\,\mu(x^{\prime\prime})\int_{x^{\prime
\prime}}^x dx^\prime\,{1\over\Pi_0}\Pi_\gamma(x^\prime),
\\
\Pi_X&=\Pi_\theta\mu + \Pi_0\left(\gamma^\prime - \left[\ln(
\Pi_\gamma^2-\tau^{\prime2})\right]^\prime\right),\\
\phi&=\sqrt{-\Pi_0}\psi,\\
\Pi_\phi&={1\over\sqrt{-\Pi_0}}\Pi_\psi,\\
{\cal Q}&={1\over\Pi_0}\int_0^{2\pi}dx\,\Bigg\{\left(\gamma-
\ln(\Pi_\gamma^2-\tau^{\prime2})\right)\Pi_\gamma\\
&\qquad-\left(\Pi_\tau+\left[\ln\left({ \Pi_\gamma-
\tau^\prime\over\tau^\prime+\Pi_\gamma}\right)\right]^\prime\right)\tau
+{1\over2}\Pi_\psi\psi\Bigg\},\\
\Pi_{\cal Q}&=\Pi_0.
\end{eqnarray}
\endnumparts 

In terms of these canonical variables the constraints
\begin{equation}
H_\perp=0,\quad H_x=0,\quad \Pi_\theta=0
\end{equation}
 are equivalent to the first-class constraints
\begin{equation}
C:=\Pi_TX^\prime + \Pi_X
T^\prime+ {1\over2}\left({1\over T}\Pi_{\phi}^2+ T\phi^{\prime2}\right)
\label{C}
=0,
\end{equation}
\begin{equation}
C_x:=\Pi_T T^\prime + \Pi_X X^\prime +\Pi_\phi\phi^\prime
=0,
\label{Cx}
\end{equation}
and
\begin{equation}
\P:=\int_0^{2\pi}dx\,
{1\over{X^{\prime2}-T^{\prime2}}}\Bigg[X^\prime\Pi_\phi\phi^\prime-{1\over 2}T^\prime\left(T^{-1}\Pi_\phi^2
+T\phi^{\prime2}
\right)
\Bigg]= 0.
\label{P}
\end{equation}

Let us make a few remarks about these new canonical variables. From \cite{RT1996} we have the following two results. First, the variables $T$ and $X$ can be interpreted as defining an embedding of a spacelike circle into $M$ in terms of harmonic coordinates based upon the Gowdy time, all relative to a fiducial flat metric. Second the constraints $C=0=C_x$ can be solved to express the variables $\Pi_T$ and $\Pi_X$ in terms of $(T,X,\phi,\Pi_\phi)$.  It follows that to construct inequivalent observables it suffices to build them as functions of $(T,X,\phi,\Pi_\phi,\Q,\Pi_\Q)$.    

 In terms of the embedding variables, the gauge-fixing described in \S3 is given by
\begin{equation}
T = -{t\over\Pi_{\cal Q}},\quad X = x + const.,
\label{gauge2}
\end{equation}
showing how the gauge picks out a family of foliations of $M$. In this gauge, and in the new variables, the constraints are equivalent to
\begin{equation}
C=0\quad\Longrightarrow\quad\Pi_T - {1\over 2} \left({\Pi_{\cal Q}\over t}\Pi_\phi^2 + {t\over \Pi_{\cal Q}}\phi^{\prime2}\right)
=0,
\end{equation}
\begin{equation}
C_x=0\quad\Longrightarrow\quad \Pi_X + \Pi_\phi\phi^\prime = 0,
\end{equation}
and
\begin{equation}
\P=0\quad\Longrightarrow\quad\int_0^{2\pi} \Pi_\phi\phi^\prime = 0.
\label{gaugeP}
\end{equation}
The gauge conditions (\ref{gauge2}) and the constraints $C=0=C_x$ form a system of second-class constraints, allowing the elimination of the variables $(T,X,\Pi_T,\Pi_X)$ from the phase space. The resulting phase space has canonical coordinates $(\Q,\phi,  \Pi_\Q, \Pi_\phi)$ --- which are still subject to the first-class constraint (\ref{gaugeP}) --- and can be identified with the phase space $\tilde\Gamma$ of \S3.  In particular we note that the phase space $\tilde\Gamma$ is a representation of the result of reducing $\Gamma^\star$ by the constraints $C=0=C_x$.

\section{The observables}

Because the transformation (4.5) is canonical, and because none of the constraint functions $C$, $C_x$, $\P$ involve  the canonical variables $(\Q,\Pi_\Q)$, it follows immediately that $(\Q,\Pi_\Q)$ are observables:

\bigskip\noindent
{\bf Proposition 1.}

{\it The variables $\Q$ and $\Pi_\Q$ satisfy}
\begin{equation}
[\Q,C] = [\Q,C_x]=[\Q,\P]=[\Pi_\Q,C] = [\Pi_\Q,C_x]= [\Pi_\Q,\P]=0.
\end{equation}
\bigskip
Note that these are ``strong'' equalities, {\it i.e.}, the Poisson brackets vanish even off of the constraint surface.

Given Proposition 1, to uncover the remaining observables it suffices to restrict attention to functions of  $(T,X,\phi,\Pi_\phi)$. We will construct the remaining observables in two steps. First, we find a suitably large set of functions on $\Gamma^\star$, expressed in the new coordinates introduced in the last section, which have vanishing Poisson brackets with the constraint functions (\ref{C})--(\ref{Cx}). We then find a suitable number of functions of {\it these} functions which also have vanishing Poisson brackets with the remaining constraint function (\ref{P}).

The results of \cite{RT1996} show the constraint functions $C$, $C_x$ generate canonical transformations mathematically equivalent to many-fingered time evolution of a scalar field $\varphi$ on $M$ satisfying
\begin{equation}
-{\partial\over\partial \T}\left(\T{\partial\varphi\over\partial\T}\right) + \T{\partial^2\varphi\over\partial \X}= 0,
\label{fieldequation}
\end{equation}
where $(\T,\X)$ are coordinates on $M$ in which the (spacelike) slices are embedded according to
\begin{equation}
\T = T(x),\quad \X = X(x).
\end{equation}
The scalar field thus has the form
\begin{equation}
\eqalign{
\varphi(\T,\X) = &{1\over\sqrt{2\pi}}(q + p \ln \T)\cr
&+{1\over2\sqrt{2}} \sum_{n=-\infty\atop n\neq 0}^\infty\left( a_n  H_0(|n|\T)e^{-in \X} + a^*_n  H^*_0(|n|\T)e^{in \X}\right),
}
\label{soln}
\end{equation}
where $H_0$ is the zeroth-order Hankel function of the second kind, satisfying
\begin{equation}
 H_0^{\prime\prime}(z) + {1\over z} H_0^\prime(z) +  H_0(z) = 0,\quad z>0.
 \label{Bessel}
\end{equation}

On a given slice $(T(x),X(x))$ the canonical variables $(\phi,\Pi_\phi)$ are 
related to the ``spacetime field'' $\varphi$ via
\begin{equation}
\phi(x) = \varphi(T(x),X(x)), 
\label{phi}
\end{equation}
\begin{equation}
\Pi_\phi(x) = T(x)\Big(X^\prime(x)\varphi_{,\T}(T(x),X(x)) + T^\prime(x)\varphi_{,\X}(T(x),X(x))\Big).
\label{piphi}
\end{equation}
The field equation (\ref{fieldequation}), being derivable from a Lagrangian, 
\begin{equation}
L= {1\over 2} T (\varphi_{,\rm T}^2 - \varphi_{,\rm X}^2),
\end{equation}
 admits a closed 1-form $\omega(f,g)$ on $M$ built from a pair of solutions $f$ and $g$ \cite{DeWitt1965}: 
\begin{equation}
\omega(f,g) = {\rm T}\left[({\partial f\over \partial T} g - {\partial g\over \partial T} f) d{\rm X} 
+ ({\partial f\over \partial X} g - f{\partial g\over \partial X}) d{\rm T}\right].
\end{equation} 
This 1-form defines a symplectic form $\Omega(f,g)$ on the space of solutions to (\ref{fieldequation}) via
\begin{equation}
\Omega(f,g) = \int_{S^1} j^*\omega,
\label{jomega}
\end{equation}
where $j\colon S^1\to M$ is any embedding. Because $\omega$ is closed (when $f$ and $g$ satisfy the field equation (\ref{fieldequation})), $\Omega$ depends at most upon the homology class of the embedding.\footnote{We will assume that the embedded circle is always in the same homology class as the non-contractible $S^1$ in $M={\bf R}^+\times S^1$.} The value of $\Omega(f,g)$ is therefore unchanged by any continuous deformation of the embedding. The constraint functions (\ref{C})--(\ref{Cx}) generate canonical transformations which deform  the embedding $(T,X)$ and give the corresponding changes in the fields $(\phi,\Pi_\phi)$ according to the field equation (\ref{fieldequation}).  The embedding independence of  $\Omega(f,g)$ can thus be used to construct phase space functions which have vanishing Poisson brackets with the constraint functions (\ref{C})--(\ref{Cx}).  The details are as follows.

We introduce a basis for the space of solutions to (\ref{fieldequation}):
\begin{equation}
u_0 = {1\over\sqrt{2\pi}} \ln \T,\quad \tilde u_0 = {1\over\sqrt{2\pi}} 
\end{equation}
\begin{equation}
u_n = {1\over2\sqrt{2}} H_0(|n| \T) e^{-in\X},\quad n\neq 0
\end{equation}
satisfying
\begin{equation}
 \Omega(u_0,\tilde u_0) = 1,
\end{equation}
\begin{equation}
\Omega(u_0, u_n) = 0 =\Omega(\tilde u_0,u_n),
\end{equation}
and
\begin{equation}
\Omega(u_n,u_m) = 0,\quad \Omega(u_n^*,u_m) = i\delta_{nm}.
\end{equation}
The coefficients in (\ref{soln}) are then given by the embedding independent formulas
\begin{equation}
q = \Omega(u_0,\varphi),\quad p =  -\Omega(\tilde u_0,\varphi),\quad
a_n =  {1\over i}\Omega(u_n^*,\varphi).
\label{coeff}
\end{equation}
Using the embedding
\begin{equation}
{\rm T} = T(x),\quad {\rm X} = X(x)
\end{equation}
 for $j$ in (\ref{jomega}), equations (\ref{coeff}), (\ref{phi}), and (\ref{piphi}) can be used to express the coefficients as functions on $\Gamma^\star$:
\begin{equation}
q =  {1\over \sqrt{2\pi}}\int_0^{2\pi}dx\,  [ \phi X^\prime  - \ln T\, \Pi_\phi],
\label{q}
\end{equation}
\begin{equation}
p  = {1\over \sqrt{2\pi}}\int_0^{2\pi} dx\, \Pi_\phi,
\label{p}
\end{equation}
\begin{equation}
\eqalign{
a_k = {1\over 2i\sqrt{2}}\int_0^{2\pi} dx\, &\Big\{ Te^{ikX} \phi\left[|k|X^\prime H_0^{*\prime}(|k|T) +ikT^\prime H_0^*(|k|T)\right]\cr
 &- \Pi_\phi H_0^*(|k|T)e^{ikX}\Big\}.
}
\label{ak}
\end{equation}
\begin{equation}
\eqalign{
a_k^* = -{1\over 2i\sqrt{2}}\int_0^{2\pi} dx\, &\Big\{ Te^{-ikX} \phi\left[|k|X^\prime H_0^{\prime}(|k|T) -ikT^\prime H_0(|k|T)\right]\cr
 &- \Pi_\phi H_0(|k|T)e^{-ikX}\Big\}.
}
\label{ak*}
\end{equation}
 We remark that the relations (\ref{q})--(\ref{ak*}) between the variables $(q,p,a_k,a^*_k)$ and the variables $(\phi, \Pi_\phi)$ is a bijection.

The embedding independence of $\Omega$ now implies that these functions on phase space have vanishing Poisson brackets with the constraint functions  (\ref{C})--(\ref{Cx}). 

\bigskip\noindent
{\bf Proposition 2.}

{\it The phase space functions $q$, $p$, $a_k$, $a_k^*$ defined in (\ref{q})--(\ref{ak}) satisfy
\begin{equation}
[q,C_\alpha] = [p,C_\alpha]=[a_k,C_\alpha]=[a_k^*,C_\alpha]=0,
\end{equation}
where $C_\alpha = (C,C_x)$.}

\proof

Proposition 2 follows from the embedding independence of $\Omega$. Alternatively, the Poisson brackets can be computed straightforwardly; the results follow using integration by parts and the Bessel equation (\ref{Bessel}) for the last two equalities.\quad \qed
\bigskip

\bigskip
At this point, Propositions 1 and 2 show we have a set of phase space functions $(\Q,\Pi_\Q, q, p, a_k, a_l^*)$ which have vanishing Poisson brackets with $C$ and $C_x$.  We remark that these functions can be viewed as defining coordinates on the phase space $\Gamma^\prime$ obtained from $\Gamma^\star$ upon reduction by the first-class constraints $C=0=C_x$.  These coordinates are canonical:

\bigskip\noindent
{\bf Proposition 3.}

{\it The non-vanishing Poisson brackets among the phase space functions $(\Q, \Pi_\Q, q, p, a_k, a_l^*)$ are:}
\begin{equation}
[\Q,\Pi_\Q] = 1,\quad [q,p]=1,\quad [a_k, a_l^*] = -i\delta_{kl}.
\end{equation}

\proof

These Poisson brackets follow from direct computation and the use of the Wronskian identity
\begin{equation}
H_0(z) H_0^{*\prime}(z)-H_0^*(z) H_0^\prime(z)   = {4i\over \pi z}.
\end{equation}
\qed
\bigskip

The phase space $\tilde\Gamma$ obtained in \S3 is a representation of $\Gamma^\prime$ associated with the choice of gauge (\ref{gauge2}). In the gauge-fixed treatment of \S3 the reduced phase space $\hat\Gamma$ was obtained from the constraint surface $\K=0$ in $\tilde\Gamma$  by taking a quotient by the group action $\K$ generates. In the gauge-invariant treatment being used in this section the corresponding procedure is to impose the constraint $\P=0$ on $\Gamma^\prime$ and take the quotient by the group action $\P$ generates. This we do as follows.

 To begin, we express $\P$ in terms of the functions $(\Q,\Pi_\Q, q, p, a_k, a_k^*)$. By direct computation we have the following.

\bigskip\noindent
{\bf Proposition 4.}

{\it In terms of the functions $(\Q,\Pi_\Q, q, p, a_k, a_k^*)$, the constraint function $\P$ in (\ref{P}) takes the form}
\begin{equation}
\P=\sum_{n=-\infty}^\infty n a_n^* a_n.
\end{equation}
\bigskip
\noindent
From Propositions 3 and 4 it follows immediately that 
$$
[q,\P]=[p,\P]=  0.
$$
Recalling Propositions 1 and 2 we now have the following result.

\bigskip\bigskip\noindent
{\bf Proposition 5.}

{\it The functions $(q,p,\Q,\Pi_\Q)$ are observables.}
\bigskip

The remaining variables, $a_k$ and $a_k^*$, $k=\pm1, \pm2, \dots$, are not observables because they are not invariant under the 1-parameter group action $G=S^1$ generated by $\P$. Given the form of $\P$ shown in Proposition 4, and using the Poisson brackets given in Proposition 3,  it is straightforward to verify these variables transform as
\begin{equation}
a_k \longrightarrow e^{-i\xi k} a_k,\quad a_k^*\longrightarrow e^{i\xi k} a_k^*,
\label{groupaction}
\end{equation}
for $\xi\in(0,2\pi)$.
Because of the simplicity of this group action it is not hard to find all the invariants, at least locally.  

 Away from the singular set $a_k=0,\ \forall\ k$, at least one of the variables must be non-vanishing, so for simplicity consider a ($G$-invariant) region of phase space $U_1$ in which $a_1\neq0$. Set $a_1=\rho e^{i\theta}$. The variables $(q, p, \Q,\Pi_\Q, \rho,\theta, a_k, a_k^*)$, $k=-1, \pm2, \pm3, \dots$, correspond to a chart $U_1^\prime$ on $\Gamma^\prime$.  We now will consider the reduction of $U_1^\prime$ by the constraint $\P=0$ which will yield a chart on $\hat\Gamma$.  On the constraint surface $\P=0$ we can eliminate $\rho$ in terms of the other variables:
 \begin{equation}
 \rho^2 = -\sum_{k\neq 1} k|a_k|^2.
 \label{rho}
 \end{equation}
Consider an observable built as a function of the variables $a_k$ and $a_k^*$, $k=\pm1, \pm2, \dots$. In light of  (\ref{rho}),  this observable is equivalent to a {\it strongly} invariant function  $F=F(\theta, a_k, a_k^*)$, $k=-1, \pm2, \pm3, \dots$. Because $G$ is connected, $G$-invariance of $F$ is equivalent to  infinitesimal invariance:
\begin{equation}
{\partial F\over \partial \theta}- i\sum_{k\neq1} k\left({\partial F\over \partial a_k}a_k - {\partial F\over \partial a^*_k}a^*_k\right) = 0.
\label{inv}
\end{equation}
Define the variables
\begin{equation}
\sigma^k = e^{ik\theta} a_k,\quad  \sigma^{*k} = e^{-ik\theta} a^*_k,\quad k\neq1.
\end{equation}
The variables $(q, p, \Q, \Pi_Q, \rho, \theta, \sigma^k, \sigma^{*k})$ also correspond to coordinates on $U_1^\prime$.
It is easy to check that the general solution to (\ref{inv}) is given by 
\begin{equation}
F= \Phi(\sigma^k, \sigma^{*k}),
\end{equation}
where $\Phi$ is any function of its arguments. 

 Using this result and Proposition 5 we have our main result.
 
 \bigskip
 \noindent
 {\bf Theorem.} 
 
 {\it On $U_1$, each observable  for the polarized Gowdy model is equivalent to a function}
 \begin{equation}
F = F(q, p, \Q,\Pi_\Q, \sigma^k, \sigma^{*k}),\quad \forall\ k\neq 1.
 \end{equation}
 
 \bigskip
 Evidently, the functions  $(q, p, \Q,\Pi_\Q, \sigma^k, \sigma^{*k})\ \forall k\neq 1$ correspond to a chart on the reduced phase space $\hat\Gamma$ for the Gowdy model.   The non-vanishing Poisson brackets for these functions are easily checked to be
 \begin{equation}
 [q,p]=1, \quad [\Q,\Pi_\Q] = 1,\quad [\sigma^k,\sigma^{*l}] = -i\delta^{kl}.
 \end{equation}
In the gauge (\ref{gauge2}) these coordinates correspond to those obtained by gauge fixing in \S3 --- up to a redefinition of $\gamma_0$ --- as can be seen by Fourier analyzing $\phi$ and $\Pi_\phi$ and then reducing by $K=0$ in an analogous fashion to what was done in this section.

\bigskip\bigskip
\ack

This work was supported in part by National Science Foundation grant PHY-0244765 to Utah State University.

\end{document}